\documentclass{aastex62}
\usepackage{newtxtext,newtxmath}
\usepackage[T1]{fontenc}
\usepackage{graphicx,amsmath,amssymb}
\usepackage{todonotes}

\newcommand{\EE}{\mathbf{E}}
\newcommand{\BB}{\mathbf{B}}
\newcommand{\II}{\mathrm{i}\,}
\renewcommand{\Re}{\mathrm{Re}\,}
\renewcommand{\Im}{\mathrm{Im}\,}

\renewcommand{\vec}{\mathbf}
\renewcommand{\oe}{{0;\mathrm{e}}}
\newcommand{\be}{{b;\mathrm{e}}}
\newcommand{\e}{\mathrm{e}}

\newcommand{\figref}[1]{Figure~\ref{#1}}

\shorttitle{The gyroresonant Bell instability}
\shortauthors{Weidl, Winske, and Niemann}

\begin{document}
\title{On the background-gyroresonant character of Bell's instability in the large-current regime}

\author[0000-0002-3440-3225]{Martin~S.~Weidl}
\affiliation{UCLA Physics \& Astronomy}

\author{Dan~Winske}
\affiliation{Los Alamos National Laboratory}

\author[0000-0001-5489-9144]{Christoph~Niemann}
\affiliation{UCLA Physics \& Astronomy}

\correspondingauthor{Martin~S.~Weidl}
\email{mwei@physics.ucla.edu}

\begin{abstract}
We show that the Bell instability, which is widely considered potentially important for cosmic-ray acceleration, is the low-frequency limit of a gyroresonant interaction between the protons of the interstellar medium and shear-Alfv\'en waves. At large cosmic-ray current densities, its growth rate is therefore limited by the proton gyrofrequency, and two modes emerge from the cold-beam dispersion relation. A third mode driven by electron gyroresonance is only weakly unstable at low current densities. We discuss implications for magnetic-field amplification and its saturation in the vicinity of supernova remnants.
\end{abstract}

%\keywords{Shock waves --- waves --- ISM: magnetic fields}

\section{Introduction}

To explain cosmic-ray (CR) acceleration to the observed knee energy $E\sim10^{15}$ eV, astrophysicists commonly consider collisionless shocks as the sites of diffusive shock acceleration or DSA \citep{axford77,bell78,blandford78}. Since the most efficient environment for DSA is a parallel collisionless shock \citep{caprioli14}, the magnetic turbulence that accelerates knee-energy CRs is likely generated by parallel beam instabilities that couple the most energetic debris of supernova remnants (SNR) with the surrounding interstellar medium. Commonly studied in this context are the gyroresonant CR streaming instability \citep{kulsrud69} and the so-called non-resonant cosmic-ray current-driven instability \citep{bell04}, widely referred to as Bell instability.

The latter mode has been investigated extensively in recent years since it grows more rapidly than the streaming instability discovered by Kulsrud, at least for sufficiently high-energy CR particles and a cold interstellar medium. In addition to its role in DSA, the Bell instability has been conjectured as an agent of magnetic-field amplification (MFA) to explain observations of high-energy electron-synchrotron radiation. Although it is sometimes referred to as a magnetohydrodynamic (MHD) instability, all of its analytical derivations depart from the kinetic Vlasov equation for a parallel-beam configuration \citep{achterberg83,bell04,amato09,zweibel10}. Early MHD simulations added a static CR current and a charge-compensating return current to Amp\`ere's law and showed significant magnetic-field growth, which saturated at values of $\delta B/B_0 \sim 10$ because of fieldline tension \citep{bell04,zirakashvili08}. \citet{pelletier06} argued that the instability would saturate because of a nonlinear kinetic effect, i.e.\ fieldline filaments growing to diameters comparable to the CR gyroradius.

Computational astrophysicists who ran fully kinetic particle-in-cell (PIC) simulations had to choose between resolving the small-wavelength Bell mode or containing all growing long-wavelength modes in the simulation domain and understandably opted for the former. Starting with a monoenergetic beam of CR quasiparticles, these simulations showed that field growth saturated at $\delta B/B_0\lesssim10$ because of a rapid decrease in the relative drift velocity: as the cosmic-ray particles decelerated, the background ions quickly accelerated \citep{niemiec08,riquelme09}. Simultaneously the dominant wavelength was growing, seemingly without bound, and the ions began to form density filaments. A deceleration of cosmic-rays was also observed by \citet{lucek00}, using a hybrid code that treated the background plasma as a MHD fluid and the cosmic-ray particles kinetically.

Later, two- and even three-dimensional hybrid simulations showed that a right-hand polarized mode dominates at lower Mach numbers and that the left-hand polarized Bell instability only grows faster for high CR currents \citep{gargate12}, and that MFA could even reach $\delta B/B_0\sim10^2$ locally within the forming density filaments \citep{caprioli13}.  However, those simulations never explained in detail the physical mechanism that simultaneously decelerated the CRs and accelerated the background ions. The reason for this background acceleration, as we will explain below, is the ultimately near-gyroresonant nature of the Bell instability.

We first sketch the derivation of parallel cold-beam instabilities in section~\ref{secTheory}. Section~\ref{secBell} reviews the approximations that Bell's original article is based on, while section~\ref{secHybrid} shows how allowing solutions with higher frequencies reveals the free-energy source of the instability: the gyromotion of background ions. In section~\ref{secElectron}, we investigate an additional background-gyroresonant mode. Astrophysical implications for SNR shock acceleration are discussed in section~\ref{secDiscussion}.

\section{Full dispersion relation}
\label{secTheory}

\subsection{Derivation sketch of the full dispersion relation}
We consider a plasma in a unidirectional magnetic field $\BB_0 = B_0\, \vec{\hat x}$ with $B_0>0$. Let the plasma consist of a background and a beam ion species, both positively charged, and a neutralizing electron population. The temperature of each species is low enough that $\beta=v_{\mathrm{th}}^2/v_A^2\ll1$, where thermal and Alfv\'en velocities are denoted by $v_\mathrm{th}$ and $v_A$, respectively.

Starting from the momentum equations for cold ions and electrons, distinguished by the index $\alpha$, and the wave equation for a current-driven EM wave,
\begin{equation}
\begin{aligned}
\mathrm D_{\alpha} \vec v_\alpha &= \Omega_\alpha\ (\EE + \vec v_\alpha/c \times \BB)\ B_0^{-1},\\
\vec j_\alpha &= q_\alpha n_\alpha \vec v_\alpha,\\
\nabla \times (\nabla \times \EE) &= -c^{-2}\, \left(\partial_t^2\, \EE + c\, \partial_t \sum_\alpha \vec j_\alpha\right),
\end{aligned}
\end{equation}
one can quickly derive the dispersion relation of parallel Alfv\'enic waves for charged-particle beams with a relative drift $V_\alpha$ parallel to the magnetic field \citep[e.g.][]{akhiezer1975}. The convective derivative is defined as $\mathrm D_{\alpha} = \partial_t + V_\alpha\, \partial_x$. The product of the charge-to-mass ratio $\Omega_\alpha/B_0$, charge $q_\alpha$, and number density $n_\alpha$ of each species yields the  species' squared plasma frequency $\omega_{p,\alpha}^2$. Performing a Fourier transform with argument $k_\parallel$ in $x$ space and a Laplace transform with argument $\II \omega$ in time (and taking $\lim \vec v_\alpha(0^+) \to V_\alpha\, \mathbf{\hat x}$), we find for the transformed perpendicular field components $\vec E_\perp = (\tilde E_y, \tilde E_z)^\top$
\begin{equation}
\left(\omega^2 - c^2\,k_\parallel^2\right)\, \vec E_\perp = \sum_\alpha \omega_{p,\alpha}^2\ \frac{\omega-V_\alpha k_\parallel}{(\omega-V_\alpha k_\parallel)^2-\Omega_\alpha^2} \left( \begin{matrix} \omega-V_\alpha k_\parallel & -\II \Omega_\alpha \\ \II \Omega_\alpha & \omega-V_\alpha k_\parallel \end{matrix} \right) \vec E_\perp.
\end{equation}
Diagonalizing this equation by introducing $E_{\mathrm{R/L}}=\tilde E_y\pm\II \tilde E_z$, one obtains
\begin{equation}
\left(\omega^2 - c^2\,k_\parallel^2\right)\, E_{\mathrm{R/L}} = \sum_\alpha \omega_{p,\alpha}^2 \frac{\omega-V_\alpha k_\parallel}{\omega-V_\alpha k_\parallel \pm \Omega_\alpha} E_{\mathrm{R/L}}.
\label{eqnErl}
\end{equation}

\begin{figure}
\centering
\includegraphics[scale=.95]{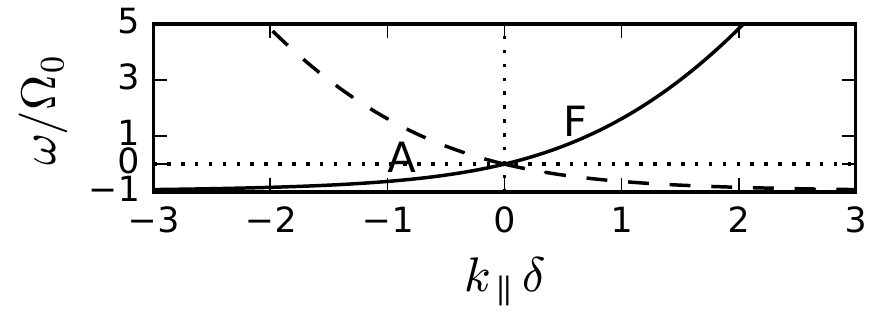}
\caption{Dispersion relation~\eqref{eqnDR} for left- and right-handed Alfv\'en waves propagating to the right (solid, $\omega/k_\parallel>0$) and to the left (dashed, $\omega/k_\parallel<0$). Frequency regimes of the rightwards fast magnetosonic (F) and shear-Alfv\'en (A) wave are indicated.}
\label{figAlfvenDisp}
\end{figure}

\subsection{Signed frequencies and other conventions}

The two dispersion relations~(\ref{eqnErl}) can be unified into one without loss of generality if one understands what the signs of frequency and wavenumber imply in a single frame of reference, following e.g.\ \citet{gary91}. We designate left-handed polarization around the magnetic-field direction, i.e.\ the polarization of the ion gyromotion, with a negative frequency. This choice, consistent with the upper sign in the denominator of (\ref{eqnErl}), corresponds to the standard Fourier sign convention with $\exp(+\II k_\parallel x - \II \omega t)$, provided that $B_0>0$. Polarization expresses in which direction the perpendicular magnetic-field vector rotates if one measures it at a fixed position in space --- `fixed' in a frame of reference that must be specified. If all ions and electrons share one common rest frame, that frame is the obvious choice. Thus we distinguish fast magnetosonic/whistler waves (with right-handed circular polarization and positive frequency for forward-propagating waves) and shear-Alfv\'en/ion-cyclotron waves (with left-handed circular polarization and negative frequency) by the sign of $\omega$. For the ion-frequency range ($\omega\ll|\Omega_e|$) and one effective ion species, the solid lines in \figref{figAlfvenDisp} show how the dispersion branches for both forward-propagating modes merge into one continuous branch with this convention, which is described by the relation \citep[e.g.][]{kulsrud2005}
\begin{equation}
k_\parallel^2 = \frac{\omega^2}{c^2}\left(1+\frac{\omega_p^2}{\Omega_0\,(\omega+\Omega_0)}\right)
\label{eqnDR}
\end{equation}
and the condition $\omega/k>0$ for rightward or forward propagation. In terms of the ion inertial length $\delta$, the reference plasma frequency $\omega_p=(n_e\,e^2/m)^{1/2}$ (in Heaviside--Lorentz units with $m$ the ion mass) is given by $\omega_p=c/\delta$. The Alfv\'en velocity $v_A = c\,\omega_p/\Omega_0$ is the phase velocity in the large-wavelength limit $k_\parallel \to 0$.

In contrast to polarization, helicity describes the direction in which a magnetic fieldline winds around the $z$ axis if one traces it in space at a fixed point in time. Helicity is determined from the sign of $k_\parallel$ that dominates in its Fourier spectrum: modes with positive helicity result in a left-handed fieldline as one follows its trajectory at a fixed point in time, while $k_\parallel<0$ corresponds to right-handed helicity (RHH). Whereas helicity is invariant under proper Lorentz transformations (which can change the norm but not the sign of $k_\parallel$) and thus well-suited for characterizing modes independently of the reference frame, polarization depends on the frame one decides or is forced to work in. Resonant coupling of a wave to the gyromotion of an ion species (as for an ion-cyclotron wave) occurs only if $\varpi' \approx -\Omega_g$, where $\varpi'$ is the angular wave frequency Doppler-shifted to the rest frame of the ion with gyrofrequency $\Omega_g$. As we will show in section~\ref{secHybrid}, gyroresonant coupling in the sense of energy transfer between a streaming ion and electromagnetic waves occurs for
\begin{equation}
|\varpi' + \Omega_g| = |\varpi - k_\parallel\, v_\parallel + \Omega_g| \leq \eta\,\Omega_g,
\label{eqnGyroresonance}
\end{equation}
where $v_\parallel$ is the parallel velocity of the ion in the same frame of reference in which $\varpi$ is the measured wave frequency and the exact value of $\eta\lesssim1$ depends on the ion densities. If the left-hand side is equal to zero, we speak of \emph{exact} gyroresonance; if it is non-zero but so small that the inequality is still fulfilled, we speak of \emph{near}-gyroresonance. The distinct phenomenon of Landau resonance ($\varpi \approx k_\parallel v_\parallel$) bears no relevance for the purely electromagnetic modes described by the dispersion relation~(\ref{eqnErl}).

Although helicity (i.e.\ $k_\parallel$) is useful to describe a mode independently of the reference frame, the frame-dependent polarization (i.e.\ $\omega'$) is not. Speaking of polarization as a property of a mode or using it to determine gyroresonance can be ambiguous unless the reference frame is clear from the context.

\section{Barely resonant: the Bell regime}
\label{secBell}

The most general configuration of two cold ion--electron plasmas counterstreaming along a magnetic field is described by the dispersion relation
\begin{widetext}
\begin{equation}
	\omega^2 - c^2\, k_\parallel^2 = \underbrace{\omega_{p,0}^2 \frac{\omega-k_\parallel v_0}{\omega-k_\parallel v_0 + \Omega_0}}_{\sigma_0} + \underbrace{\omega_{p,\oe}^2 \frac{\omega-k_\parallel v_\oe}{\omega - k_\parallel v_\oe +\Omega_{\e}}}_{\sigma_\oe} + \underbrace{\omega_{p,\be}^2 \frac{\omega-k_\parallel v_\be}{\omega - k_\parallel v_\be +\Omega_{\e}}}_{\sigma_\be} + \underbrace{\omega_{p,b}^2 \frac{\omega-k_\parallel v_b}{\omega-k_\parallel v_b + \Omega_b}}_{\sigma_b},
\label{eqnTotalDR}
\end{equation}
\end{widetext}
with the background and beam ion species denoted by the subscripts $0$ and $b$, respectively, and $e$ denoting the corresponding electron species. As noted above, admitting negative $\omega$ and $k_\parallel$ allows us to consider only one of the signs in (\ref{eqnDR}). The background electron species compensates the background-ion current, i.e.\ $n_\oe=n_0$ and $v_\oe=v_0$, and the beam electrons compensate the beam-ion current likewise. 

The intent of this article is to provide a systematic analysis of this dispersion relation in various limits, in order to compare how one and the same instability can be interpreted respectively by astrophysicists, who often have an MHD model in mind, space physicists, who generally think in terms of ion kinetics, and basic plasma physicists, who are occasionally forced to consider electron kinetics. In this section, we clarify some of the assumptions made by \citet{bell04} to derive an apparently non-resonant instability from an electron current drifting in a MHD plasma. Our goal is the simplest growing-mode analysis possible in the low-frequency range --- if one measures $\omega$ in the background frame. We can thus fix our frame of reference by setting $v_0\equiv v_\oe\equiv0$; the beam velocity in this frame is $V_b$.

As a first approximation, we assume that $|\Omega_e|$ is much larger than the frequency of any growing modes, not only in the background frame ($|\omega/\Omega_e|\ll1$) but also in the beam frame ($|\omega-k_\parallel V_b|\ll|\Omega_e|$). Expanding $\sigma_\oe$ and $\sigma_\be$ to first order in these frequencies (the `hybrid' approximation) and using $\omega_{p,0}^2\,\Omega_e = -\omega_{p,\oe}^2\,\Omega_0$, we obtain
\begin{equation}
\sigma_\oe + \sigma_\be = -\ \omega_{p,0}^2\, \frac{\omega}{\Omega_0} - \omega_{p,b}^2\, \frac{\omega-k_\parallel V_b}{\Omega_b} + \mathcal O \left(\frac{\omega^2}{\Omega_e^2}\right).
\label{eqnHybridApprox}
\end{equation}

As a second approximation, we limit our analysis to the magnetohydrodynamic frequency range, where $|\omega| \ll \Omega_0$ in the background frame. Applying this `semi-MHD' expansion to the background-ion contribution only yields
\begin{equation}
\sigma_0 = \omega_{p,0}^2 \left[ \frac{\omega}{\Omega_0} - \frac{\omega^2}{\Omega_0^2} + \mathcal O \left( \frac{\omega^3}{\Omega_0^3} \right) \right].
\label{eqnsMHDApprox}
\end{equation}

To these orders, the total dispersion relation can thus be written as
\begin{equation}
\omega^2 \left(1+\frac{\eta^2 c^2}{v_A^2}\right) - c^2\, k_\parallel^2 = - \omega_{p,b}^2\, \frac{\omega-k_\parallel V_b}{\Omega_b} + \sigma_b,
\end{equation}
where $\eta^2 = n_0/(n_0+n_b)$ and $v_A$ is still defined using the \emph{total} ion-mass density $B_0\, m_0^{-1/2}(n_0+n_b)^{-1/2}$, or
\begin{equation}
\omega^2 \left(1+\frac{\eta^2 c^2}{v_A^2}\right) + \frac{\eta^2 c^2}{v_A^2}\, k_\parallel B_0\, \frac{e\,n_b\,V_b}{m\,n_0}\, \left(\frac{\omega}{k_\parallel V_b}-1\right) - c^2\, k_\parallel^2 = \sigma_b.
\label{eqnBell}
\end{equation}

\subsection{Bell's standing-waves limit}
The simplest version of Bell's instability is found by limiting the search range to slowly propagating short-wavelength modes far from Landau resonance with the beam, such that $|\omega|\ll |k_\parallel V_b|$ and the term linear in $\omega$ can be dropped. Imposing furthermore that the frequency of the Bell mode in the beam frame is too small for exact gyroresonance with the beam ions, such that $|\omega-k_\parallel V_b|\ll\Omega_b$, we set $\sigma_b\approx0$. Neither assumption is necessary to solve the cubic polynomial~(\ref{eqnBell}) for $\omega$, but these simplifications result in the straightforward equation
\begin{equation}
\omega^2 = \frac{v_A^2\,c^2}{v_A^2 + \eta^2 c^2} \left( k_\parallel^2 + \eta^2 B_0\, \frac{j_b}{m_0 n_0 v_A^2}\, k_\parallel \right).
\label{eqnSWL}
\end{equation}
We thus obtain a band of growing modes in the range $-\eta^2 B_0 j_b/(m_0n_0v_A^2)<k_\parallel<0$, with the beam current $j_b=e\,n_b\,V_b$. The peak growth rate
\begin{equation}
\gamma_{\max} = \frac{v_A}2\, \frac{c}{\sqrt{v_A^2 + \eta^2 c^2}}\, \eta^2 B_0\, \frac{j_b}{m_0n_0v_A^2}
\label{eqnGBell}
\end{equation}
occurs at
\begin{equation}
k_{\max} = - \frac12\, \frac{\eta^2\, B_0\, j_b}{m_0\,n_0\,v_A^2}.
\label{eqnKBell}
\end{equation}

In the limits $v_A\ll c$ and $n_b \ll n_0$, this is exactly identical to the cosmic-ray current-driven instability of \citet{bell04}, who used the opposite sign convention and thus found a positive $k_{\max}$. Since we dropped the beam-ion term $\sigma_b$, it is clear that this mode must result from an interaction between the background plasma (represented by $\sigma_0+\sigma_\oe$) and the beam electrons ($\sigma_\be$). The negative helicity ($k_{\max}<0$) implies that, in a snapshot of the waves driven by this instability, any magnetic fieldline forms a right-handed helix around the $x$~axis. Were one to observe the evolution of these waves from a point fixed in the beam frame, one would measure them as right-hand polarized (since $\Re\omega-k_{\max} V_b>0$), but in the background frame they appear to be purely growing and hence completely unpolarized ($\Re\omega=0$) in the two non-beam-resonant limits that we have applied.

\subsection{Uncompensated electron current}
If we relent on the condition $|\omega|\ll |k_\parallel V_b|$ but continue ignoring any gyroresonant interaction with the beam ions, we effectively demand that the left-hand side of equation~(\ref{eqnBell}), now including the $\mathcal O(\omega/k_\parallel V_b)$ term, be equal to zero. We then find that, in order for the discriminant of the quadratic polynomial in $\omega$ to become negative, the beam drift relative to the background ions must exceed \citep[cf.][]{zweibel10}
\begin{equation}
V_{b,\min} = \frac{c}{\sqrt{v_A^2 + \eta^2\, c^2}}\,v_A \longrightarrow \sqrt{\frac{n_0+n_b}{n_0}} v_A
\label{eqnVminBell}
\end{equation}
so that a band of (in the background frame) left-hand polarized waves with a uniform real angular frequency $\varpi = \Re\omega$ can grow:
\begin{equation}
\varpi = -\, \frac12\, \frac{\eta^2 c^2}{v_A^2 + \eta^2 c^2}\, \frac{n_b}{n_0}\, \Omega_0 \longrightarrow -\,\frac12\,\frac{n_b}{n_0}\, \Omega_0.
\label{eqnWBell}
\end{equation}

Let us insert this solution, derived in the semi-MHD limit, into the condition~(\ref{eqnGyroresonance}) for near-gyroresonance. The left-hand side evaluates in the background frame as $\varpi + \Omega_0 = [1-n_b/(2n_0)]\,\Omega_0$. Expanding our definition of $\eta$ to first order in $n_b$, we find that the right-hand side of (\ref{eqnGyroresonance}) is equally $\eta\,\Omega_0\approx[1-n_b/(2n_0)]\,\Omega_0$. This means that keeping only the lowest-order non-vanishing term ($\sim\omega^2/\Omega_0^2$) of the background-plasma expression $\sigma_0+\sigma_\oe$ yields waves with the smallest possible value of $|\varpi|$ that is still compatible with near-gyroresonance with the background ions. We will come back to this apparent coincidence in section~\ref{secHybrid}.

\subsection{`Complete' semi-MHD solution}
An even more accurate analysis can be performed by retaining $\sigma_b$ and solving the cubic polynomial~(\ref{eqnBell}) exactly. First, let us recall the approximations required to arrive at that cubic: we treated the background ions and electrons as a quasi-neutral plasma composed of particles with infinitely large gyrofrequency, the beam electrons as a current of negative particles with infinitely large gyrofrequency, and the beam ions as a free-streaming ion population with zero temperature and finite gyrofrequency. We have dropped terms of $\mathcal O(\omega^3/\Omega_0^3)$ and $\mathcal O(\omega^2/\Omega_e^2)$ and have kept all orders in $\omega/\Omega_b$.

\begin{figure}
\centering
\includegraphics[scale=.95]{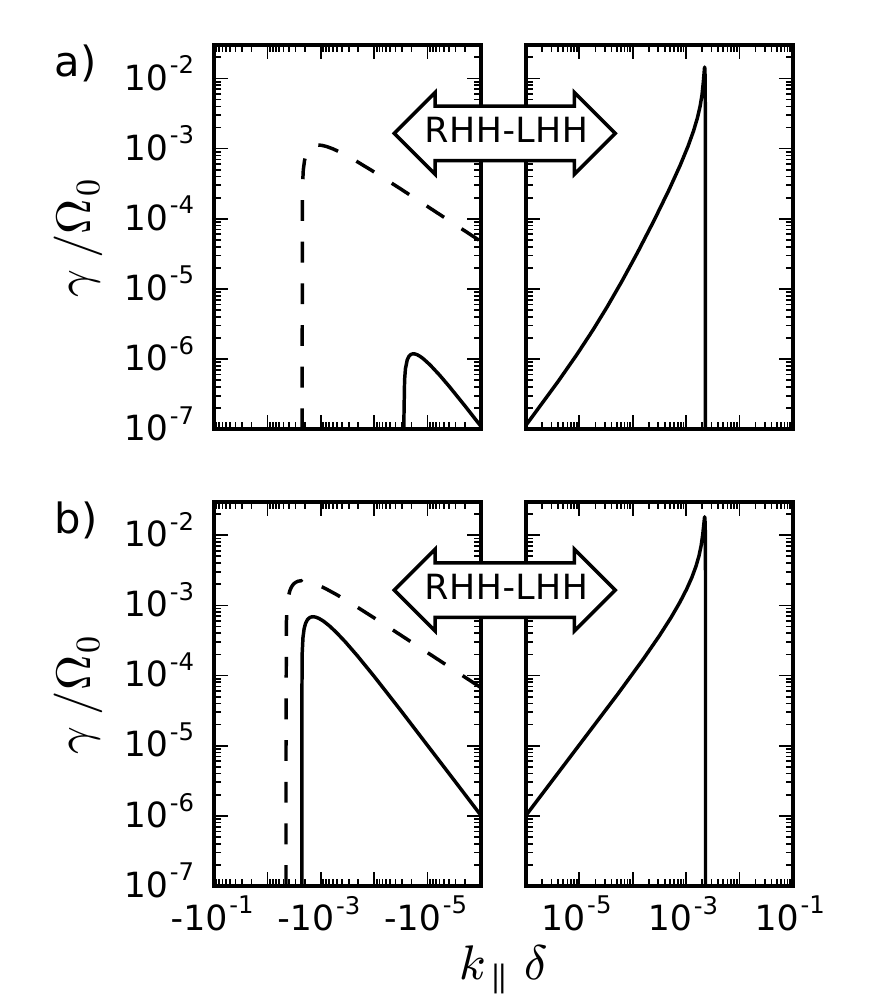}
\caption{Exact growth rates in the semi-MHD model~(\ref{eqnBell}) (solid lines) and the standing-waves limit~(\ref{eqnSWL}) (dashed lines) for negative-helicity waves, which form fieldlines that trace a right-handed helix (RHH) in space, and positive-helicity waves (LHH). The beam drift is $V_b=450\,v_A$, the beam density is a) $n_b=5\cdot10^{-6}\,n_0$, b) $n_b=10^{-5}\,n_0$.}
\label{figLowdenBell}
\end{figure}

\figref{figLowdenBell} compares the exact solution for $\gamma = \Im\omega$ of the semi-MHD model~(\ref{eqnBell}) with the analytic solution of the standing-waves limit~(\ref{eqnSWL}), in which $\sigma_b$ and thus all $\mathcal O(\omega/\Omega_b)$ terms are neglected, for a beam-background drift velocity $V_b=450\,v_A$, a typical value for supernova-remnant (SNR) scenarios, and two possible beam densities (and $c=10^4\,v_A$). The most obvious difference between the two models is the complete absence of growing waves with $k_\parallel>0$ if the beam ions are ignored. In the semi-MHD approximation these waves have a positive real frequency in the background frame; we classify them as fast magnetosonic waves (F in \figref{figAlfvenDisp}) as opposed to the negative-$\varpi$ and negative-$k_\parallel$ shear-Alfv\'en waves which are found in both models. A second difference is how sensitively the peak growth rate of these shear-Alfv\'en waves depends on the beam density: whereas doubling $n_b/n_0$ from $5\cdot10^{-6}$ to $10^{-5}$ also doubles the prediction of the standing-waves limit, the semi-MHD maximum growth rate for $k_\parallel<0$ increases by almost three orders of magnitude.

\begin{figure}
\centering
\includegraphics[scale=.95]{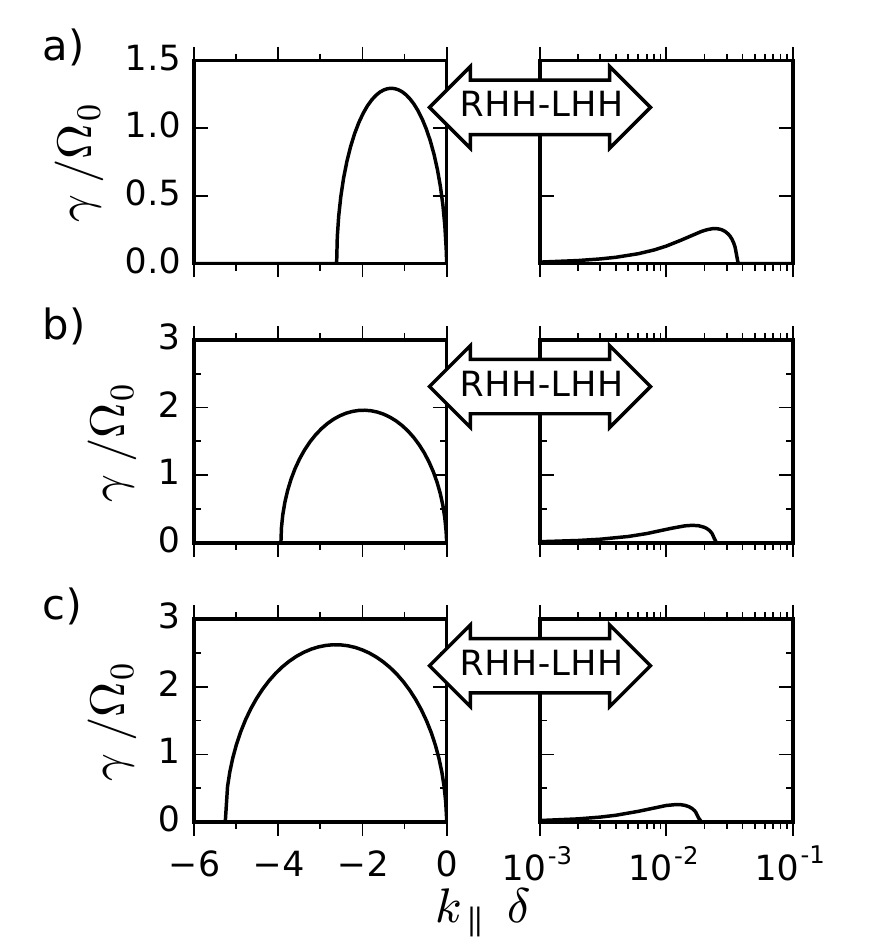}
\caption{Growth rates in the semi-MHD model~(\ref{eqnBell}) for $n_b=0.05\,n_0$ and a) $V_b=50\,v_A$, b) $75\,v_A$, c) $100\,v_A$. This model is invalid where $\gamma\gtrsim\Omega_0$.}
\label{figMeddenBell}
\end{figure}

Using an even larger beam density $n_b=0.05\,n_0$, we show the growth rates derived from the semi-MHD model for three different values of $V_b$ in \figref{figMeddenBell}. At this beam density, the solutions for $\gamma$ in the standing-waves limit are indistinguishable from the semi-MHD results for shear-Alfv\'en waves; fast magnetosonic waves are still not growing if $\sigma_b$ is neglected. The problem that arises in this regime is that ignoring $\mathcal O(\omega^3/\Omega_0^3)$ terms cannot be justified if $\gamma=\Im\omega\gtrsim\Omega_0$. Hence the predictions of the semi-MHD model compare rather unfavorably with reality, or with any simulation that resolves finite-Larmor-radius effects, as soon as $|\omega|^3/\Omega_0^3\geq1$, even though $|\varpi|/\Omega_0$ may still be negligible. From equation~(\ref{eqnGBell}), we find that the semi-MHD approximation is only valid if
\begin{equation}
j_b \ll 2\, \frac{\sqrt{v_A^2+\eta^2c^2}}{\eta^2\,c}\,e\,n_0\,v_A.
\label{eqnJbell}
\end{equation}
Together with condition~(\ref{eqnVminBell}) and the assumption $c \gg v_A$, this implies that the `Bell regime' is limited to the current range
\begin{equation}
\frac{n_b}{n_0}\, \sqrt{1+\frac{n_b}{n_0}} \ll \frac{n_b}{n_0}\,\frac{V_b}{v_A} \ll 2\, \sqrt{1+\frac{n_b}{n_0}}.
\end{equation}

\section{Background-gyroresonant: beyond bifurcation}
\label{secHybrid}

If the beam current is too strong for the semi-MHD approximation, the first term in the expansion~(\ref{eqnsMHDApprox}) can no longer cancel the hybrid expansion of $\sigma_\oe$. Assuming that we still limit our analysis to modes with $|\omega|\ll|\Omega_e|$, we can then combine the expansions of both the background and the beam electron term:
\begin{equation}
\sigma_\oe + \sigma_\be = -\,\omega_p^2\, \frac{\omega - k_\parallel V_e}{\Omega_0} + \mathcal O\left(\frac{\omega^2}{\Omega_e^2}\right),
\end{equation}
with $n_e V_e = n_b V_b$. Since $V_e$ is the average drift velocity of all electrons in the background frame, the details of the electron distribution functions do not matter in the hybrid approximation \citep[cf.][]{amato09}. Inserting this into~(\ref{eqnTotalDR}), we find for the dispersion relation in the background frame:
\begin{equation}
	\left( \omega^2 - k_\parallel^2~c^2 \right)~\frac{n_e}{\omega_p^2} =\\
	n_0~\frac{\omega}{\omega + \Omega_0} - n_e~\frac{\omega - V_e~k_\parallel}{\Omega_0} + n_b~\frac{\omega - V_b~k_\parallel}{\omega-V_b~k_\parallel+\Omega_b}.
\label{eqnHybridDR}
\end{equation}
Even if we keep the first term on the left-hand side, this is a quartic polynomial in $\omega$, for which an analytical solution can be found. For non-zero temperatures, \citet{winske84} discussed the numerical solution of the growing right-hand-helicity mode at $-1\lesssim k_\parallel \delta < 0$ that this dispersion relation predicts in detail. \citet{sentman81} presented a more general overview of growing modes.

\subsection{Hybrid-electron current}

To get initial insight into the structure of this solution, however, we momentarily ignore both the first term on the left and the last term on the right of equation~(\ref{eqnHybridDR}), assuming that $|\omega|^2 \ll \omega_p^2$  and that we can qualitatively understand the behavior of the solution in the shear-Alfv\'en range without considering the beam-ion term. The latter assumption implies that we are investigating the effects of an uncompensated electron current. It turned out to be sufficiently accurate in the Bell regime, and it will serve us well here.

Writing $\omega = \varpi + \II \gamma$, the dispersion relation becomes
\begin{equation}
\frac{\varpi + \II \gamma - k_\parallel V_e}{\Omega_0} - \frac{k_\parallel^2\, c^2}{\omega_p^2} = \eta^2\, \frac{\varpi + \II \gamma}{\varpi + \II \gamma + \Omega_0}.
\label{eqnHECdr}
\end{equation}

The imaginary part immediately yields our definition of near-gyroresonance,
\begin{equation}
\gamma\, \left( \left| \varpi + \Omega_0 \right|^2 + \gamma^2 \right) = \gamma\, \eta^2\,\Omega_0^2.
\label{eqnHECgyrores}
\end{equation}
Any solution with $\gamma\neq0$ must fulfill condition~(\ref{eqnGyroresonance}). Inserting this into the real part, we find
\begin{equation}
\varpi = \frac{\Omega_0}{2}\, \left( \delta^2\, k_\parallel^2 + \frac{V_e}{\Omega_0}\, k_\parallel - \frac{n_b}{n_0+n_b} \right).
\label{eqnHECfreq}
\end{equation}
The semi-MHD limit~(\ref{eqnWBell}) can be retrieved by dropping terms of $\mathcal O(n_b^2/n_0^2)$, which is easily justified in SNR applications, and dropping terms of $\mathcal O(\delta^2 k_\parallel^2)$ and $\mathcal O(k_\parallel V_e/\Omega_0)$.

\begin{figure}
\centering
\includegraphics[scale=.95]{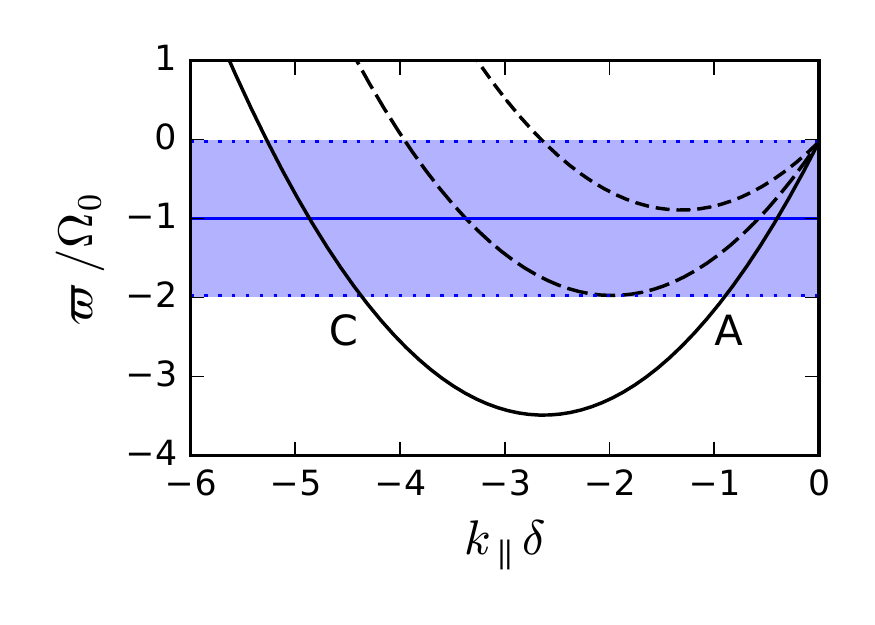}
\caption{Near-gyroresonance condition~(\ref{eqnHECgyrores}) (shaded area) and real frequency~(\ref{eqnHECfreq}) according to the hybrid-electron-current model for $n_b=0.05\,n_0$ and $V_e = V_b\, n_b/n_e $ with $V_b = 50\,v_A$ (short-dashed), $75\,v_A$ (long-dashed), and $100\,v_A$ (solid).}
\label{figGRinterv}
\end{figure}

\figref{figGRinterv} shows a graphical solution of the real part of dispersion relation~(\ref{eqnHECdr}). Assuming an uncompensated hybrid-electron current, shear-Alfv\'en waves with wavenumber $k_\parallel$ become unstable due to near-gyroresonance ($0<\gamma<\eta\,\Omega_0$) where the corresponding parabola lies within the shaded region, which represents the space where $|\varpi+\Omega_0|\leq\eta\,\Omega_0$. Exact gyroresonance and the maximum growth rate occur only where $\varpi=-\Omega_0$.

\subsection{`Complete' hybrid solution}

\begin{figure}
\centering
\includegraphics[scale=.95]{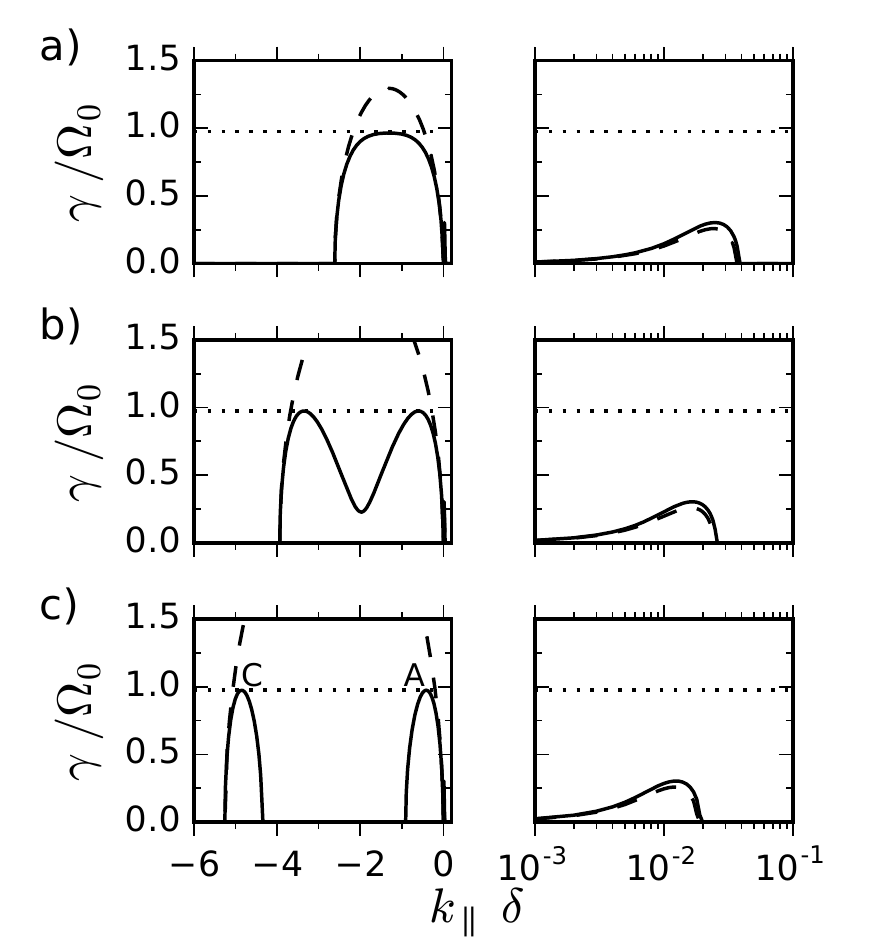}
\caption{Exact growth rates in the semi-MHD model~(\ref{eqnBell}) (dashed)} and the hybrid model~(\ref{eqnHybridDR}) (solid) for $n_b=0.05\,n_e$ and a)~$V_b = 50\,v_A$, b) $75\,v_A$, c) $100\,v_A$.
\label{figMeddenBell2}
\end{figure}

Returning to the full hybrid-model dispersion relation~(\ref{eqnHybridDR}), we compare its exact solution with the semi-MHD model in \figref{figMeddenBell2}, using parameters similar to those in \figref{figMeddenBell}. Physically, the parameter sets are not completely equivalent to those used above for the semi-MHD case --- beam density and velocity are normalized relative to the electrons now as opposed to the background ions --- but analytically this choice keeps the two models as comparable as possible. As we explained in the previous section, both models agree with each other where $|\varpi|\ll\Omega_0$ in the background frame. Moreover, the growth rate of fast magnetosonic waves with $k_\parallel>0$ varies only weakly between both models. After all, the gyroresonant behavior of the beam term $\sigma_b$, which drives these modes, is kept even in the semi-MHD model.

Yet at the points where $\varpi=-\Omega_0$ in \figref{figGRinterv}, the growth rate of shear-Alfv\'en waves reaches its maximum value $\gamma\approx\eta\Omega_0$ in the hybrid model. Where $\varpi<-\Omega_0$, the growth rate decreases again. As a result, the hybrid model allows the growing wave with $k_\parallel<0$ to split into two distinct modes if the current strength exceeds the validity limit~(\ref{eqnJbell}) of the semi-MHD approximation. To distinguish both RHH modes, we refer to the long-wavelength peak ($-1\lesssim k_\parallel\,\delta<0$) as shear-Alfv\'en wave (A) and to the short-wavelength peak as ion-cyclotron wave (C). Whether the two peaks are connected by a band of unstable modes (as in \figref{figMeddenBell2}b) or separated by a band of stable modes (as in \figref{figMeddenBell2}c) depends approximately on whether (\ref{eqnHECfreq}) yields a solution with $\varpi<-(1+\eta)\Omega_0$ (see \figref{figGRinterv}). Using the hybrid-electron-current model~(\ref{eqnHECfreq}), we find as the condition for mode-splitting
\begin{equation}
V_e > 2\,v_A\,\sqrt{2-\frac{n_b}{n_0+n_b}}\longrightarrow\sqrt8\,v_A,
\end{equation}
and as the condition for a stable band between the two modes
\begin{equation}
V_e > 2\,v_A\,\sqrt{ 2\,\left[1+\left(\frac{n_0}{n_e}\right)^{1/2}\right] - \frac{n_b}{n_e}}\longrightarrow4\,v_A.
\end{equation}
The respective conditions on the beam-ion velocity in the background frame can be obtained from $n_b\,V_b=n_e\,V_e$.

\begin{figure}
\centering
\includegraphics{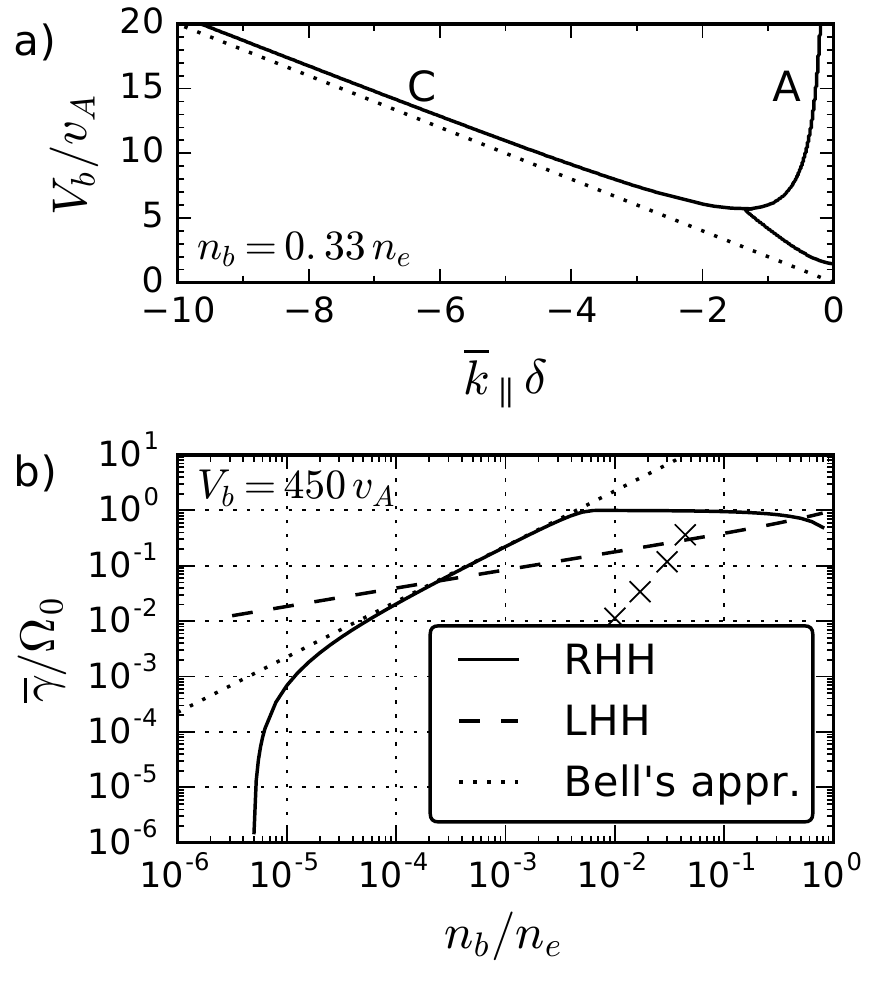}
\caption{a) Wavenumber $\overline k_\parallel$ of fastest-growing mode(s) with $k_\parallel<0$ for $n_b=0.33\,n_e$, derived from the exact solution (solid) of the hybrid-model dispersion relation (\ref{eqnHybridDR}) and from Bell's standing-waves limit (\ref{eqnKBell}) (dotted). Both the shear-Alfv\'en (A) and the ion-cyclotron (C) branches are shown. b) Fastest growth rate $\overline \gamma$ of modes forming magnetic fieldlines with right-handed helicity (RHH, shear-Alfv\'en branch, solid) and left-handed helicity (LHH, fast magnetosonic branch, dashed), respectively, for a relative beam drift $V_b=450\,v_A$ with respect to the electron rest frame; with Bell's approximation~(\ref{eqnGBell}) indicated by the dotted line. Crosses mark the growth rate of the electron-gyroresonant mode described in section~\ref{secElectron}.}
\label{figKGlog}
\end{figure}

\figref{figKGlog}a shows that this mode-splitting behavior persists at least qualitatively for a strong beam density $n_b=0.33\,n_e$, for which neglecting the beam-ion term as in the previous subsection is  quantitatively no longer appropriate. The wavenumber $\overline k_\parallel$ of the fastest-growing mode, derived from the exact solution of (\ref{eqnHybridDR}), converges towards the approximation (\ref{eqnKBell}) obtained from Bell's standing-waves limit for large beam velocities. However, this approximation becomes inaccurate in the vicinity of the bifurcation point and naturally fails to predict the existence of the long-wavelength A mode altogether.

In \figref{figKGlog}b, we compare Bell's prediction~(\ref{eqnGBell}) for the maximum growth rate $\gamma_{\mathrm{max}}$ with the exact solution $\overline\gamma$ of (\ref{eqnHybridDR}), plotting both the fastest-growing wave with $k_\parallel<0$ (RHH) and the fastest-growing fast magnetosonic wave with $k_\parallel>0$ (LHH). Bell's approximation describes the growth of shear-Alfv\'en waves well in an intermediate beam-density regime. We have already discussed why it fails at high beam densities to predict the upper limit of $\overline\gamma$ at $\Omega_0$, but it equally overestimates the growth rate at very low beam densities. This behavior was already apparent in \figref{figLowdenBell}a and is an artifact of ignoring the beam-ion term $\sigma_b$ to arrive at (\ref{eqnGBell}) --- paradoxically, the presence of beam ions suppresses the growth of shear-Alfv\'en waves by a larger relative amount at a lower relative beam density.

\newpage

\section{Electron-gyroresonant: `electromagnetic Buneman'}
\label{secElectron}

Similarly to how a second mode driven by the electron/ background-ion interaction appears if one retains the gyroresonant denominator of the background-ion term~$\sigma_0$, a third peak due to this interaction appears if one forgoes the $|\omega/\Omega_e|\ll1$ expansion of $\sigma_\oe+\sigma_\be$. This peak occurs at extremely short wavelengths and is in some sense the electromagnetic equivalent of the electrostatic electron-current-driven instability \citep{buneman58}. Nevertheless, since the mode it represents is gyroresonant with both the electrons and the background ions, it is important to consider as a possible source of magnetic field amplification and pitch-angle scattering.

We assume that all electrons drift with the uniform velocity $v_e$ and write (\ref{eqnTotalDR}) as
\begin{equation}
	\omega^2 - c^2\, k_\parallel^2 = \omega_{p,0}^2 \frac{\omega-k_\parallel v_0}{\omega-k_\parallel v_0 + \Omega_0} \; + \omega_{p,e}^2 \frac{\omega-k_\parallel v_e}{\omega - k_\parallel v_e +\Omega_{\e}} + \omega_{p,b}^2 \frac{\omega-k_\parallel v_b}{\omega-k_\parallel v_b + \Omega_b}.
\label{eqnBgElDR}
\end{equation}

As a fifth-order polynomial in $\omega$, this dispersion relation is no longer analytically solvable in closed form. However, solutions for $\omega$ can be found very efficiently by any numerical root-finding algorithm, if one knows where to search. A first rough guess $\tilde k_\parallel$ for the wavenumber we are looking for can be obtained by setting the real parts of the gyroresonant denominators of electrons and background ions to zero simultaneously, yielding
\begin{equation}
\tilde k_\parallel = \frac{\Omega_0-\Omega_e}{v_0-v_e}.
\end{equation}

\begin{figure}
\centering
\includegraphics{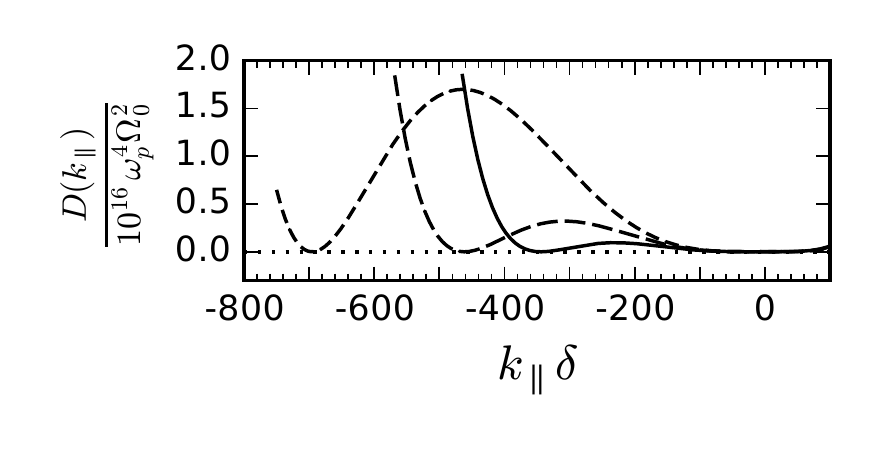}
\caption{Finding the electron-gyroresonant roots of the discriminant~(\ref{eqnDiscrim}) for $n_b=0.05\,n_0$ and $V_b/v_A=50$ (short-dashed), 75 (long-dashed), 100 (solid).}
\label{figDiscrim}
\end{figure}

A more accurate estimate results from dropping the first and last terms of (\ref{eqnBgElDR}). Ignoring the beam ions is easily justified, since a gyroresonant interaction between electrons and background ions will be far from gyroresonance with the beam ions. The leading $\omega^2$ can be discarded safely as long as $|\omega|\ll\omega_{p,0}$, which is generally true of the mode we will find. The remaining quadratic in $\omega$ predicts resonant growth where its discriminant $D$ is negative:
\begin{widetext}
\begin{equation}
D = \left[ \omega_{p,0}^2 (k V_e + \Omega_e) + \omega_{p,e}^2 (k V_e + \Omega_0) + k^2 c^2 (k V_e + \Omega_e +\Omega_0) \right]^2 - 4\, \Omega_e k (k^2 c^2 + \omega_{p,0}^2 + \omega_{p,e}^2) \left(k^2 c^2 V_e + k c^2 \Omega_0 + \omega_{p,0}^2 V_e \right).
\label{eqnDiscrim}
\end{equation}
\end{widetext}
Here $V_e=n_b V_b/n_e$ is again the drift velocity of the uniform electron population in the background-ion rest frame. It is trivial to find the minima of a sixth-order polynomial in $k_\parallel$ and check if $D$ is negative there. \figref{figDiscrim} demonstrates that, in addition to the A and C modes shown in \figref{figMeddenBell2}, an electron-resonant wave with very short wavelength ($k_\parallel\delta<-300$) becomes unstable in each of the three depicted cases.

\begin{figure}
\centering
\includegraphics{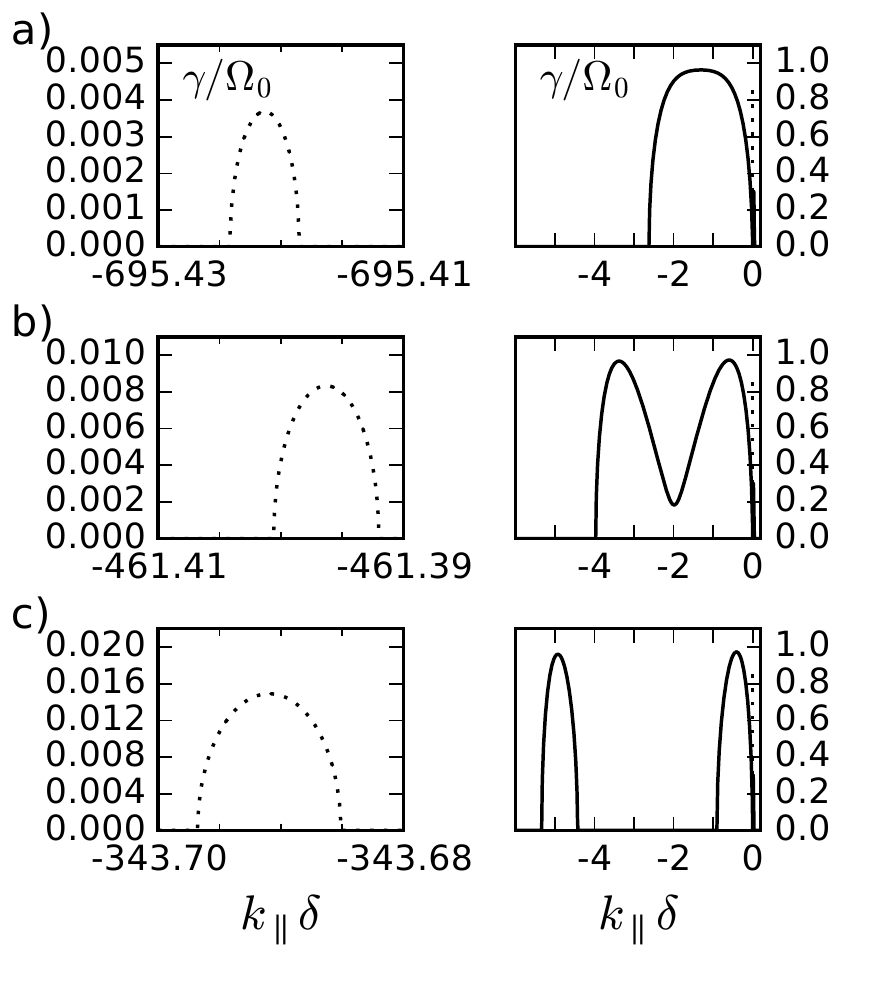}
\caption{Numerically obtained growth rates of the electron-gyroresonant mode (left) and the long-wavelength mode(s) with $k_\parallel<0$ (right) for $n_b=0.05\,n_0$ and a) $V_b=50\,v_A$, b) $V_b=75\,v_A$, c) $V_b=100\,v_A$.}
\label{figMeddenBell3}
\end{figure}

The full numerical solution of (\ref{eqnBgElDR}) for those three cases is shown in \figref{figMeddenBell3}. Although extremely narrow bands of unstable wavenumbers exist in each case at the values of $\tilde k_\parallel$ we have found above, their growth rates are only fractions of the long-wavelength modes derived in previous sections.

So at least under our idealized assumption of a cold, current-compensating electron beam, electron gyroresonance can be neglected and the hybrid model is sufficiently accurate for predicting magnetic-field growth. This holds particularly true for the low beam densities expected in SNR environments, as the insignificant growth rate of the electron-gyroresonant mode in \figref{figKGlog}b (marked by crosses) demonstrates. At a beam density of $n_b\approx0.05\,n_0$ the electron-gyroresonant mode merges with the C mode.

\section{Discussion}
\label{secDiscussion}

In several of our calculations we have used an Alfv\'enic Mach number of 450 for cosmic-ray protons escaping into the interstellar medium around a SNR. This estimate is based on an ISM magnetic-field strength of $10\,\mu\mathrm{G}$ and a background proton density of $1\,\mathrm{cm}^{-3}$. The average cosmic-ray velocity is taken to be $10^7\,\mathrm{m~s}^{-1}$, a typical speed of SNR shocks.

Crucially for the analytic treatment, we have assumed that the thermal velocities of all species are much smaller than the Alfv\'en velocity. For the background protons, this is a reasonable enough assumption; we focus on early times long before the bulk expelled in the supernova explosion arrives at and perturbs our patch of interstellar medium. For the cosmic-ray beam, a low temperature can be justified on similar grounds. As long as magnetic-field amplification is still low enough for linear theory to apply, the beam protons will be mainly free-streaming and their velocity will be simply the quotient of the distance to the SNR and the arrival time. Thus the beam velocity may initially be much higher than what we have assumed, but it will decrease over time.

We do not make any claim about the relative CR density in the vicinity of SNRs, especially because we expect this density to vary dramatically over time. As \figref{figKGlog}b shows, the growth rate of right-hand-helicity modes rolls off rapidly at CR densities below $10^{-5}\,n_0$, so Bell's instability is unlikely to contribute to strong field amplification in that regime. We have noted that this roll-off results from the beam-ion term $\sigma_b$; hence a non-zero-temperature CR distribution may mitigate the roll-off and shift it to lower densities. The same caveat applies to the dominance of left-hand-helicity modes (i.e.\ the CR streaming instability), which occurs for densities of (cold) CR protons below $3\cdot10^{-4}\,n_0$. At this density, the growth rate of both the left- and the right-hand polarized mode is $0.05\,\Omega_0$; the corresponding $e$-folding time is about three minutes.

If one assumes an extremely large beam temperature \citep[as in][]{bell04}, a fraction of the CR protons has a velocity smaller than $V_e$ in the background frame. This scenario corresponds to the resonant left-hand instability (LHI) discussed by \citet{gary84}, as these CR protons can gyroresonantly amplify right-hand helicity waves (or left-hand polarized waves in the background frame), functioning similarly to the background ions in our model. The growth rate of these waves in Bell's original set-up is therefore slightly larger than in our cold-beam configuration. However, since the LHI is only driven by the far left tail end of the CR distribution, it grows extremely slowly even at large beam densities. In the far upstream region of SNR shocks, it is almost certainly insignificant.

Electrons have been assumed to be cold and current-compensating. The former assumption is easily defended: for the modes that we have shown to dominate, the electron temperature is irrelevant. Both the A and C modes derived from the hybrid model~(\ref{eqnHybridDR}) are independent of second-or-higher moments of the electron distribution function. The growth of the electron-gyroresonant mode obviously does depend on the electron temperature, but its growth rate is so low to begin with that electron heating is unlikely to make this mode competitive in an SNR context. However, in our analysis we have focused on electromagnetic modes and ignored oblique electrostatic modes, particularly ion-acoustic modes driven by Landau damping and the (actual) Buneman instability. A hot electron distribution may promote their growth into a regime where nonlinear interactions with the electromagnetic modes become important \citep[see][]{wiener13}.

Our latter assumption about electrons, i.e.\ that they compensate the electrical current represented by the escaping CR protons, will be accurate eventually because of Lenz's law. But the escaping cosmic-ray protons will almost certainly be preceded by a burst of fast electrons accelerated in the supernova explosion, and the effect of this burst on the electron population of the surrounding interstellar medium depends on a large number of variables and is difficult to predict in general. If the resulting drift speed between electrons and the background protons is much larger than what we have assumed above, the electron-gyroresonant mode could potentially amplify the magnetic field on short length scales much faster.

\section{Summary}

We have found that the `non-resonant cosmic-ray current-driven' instability, or Bell instability, is driven by a nearly gyroresonant interaction between the background protons and shear-Alfv\'en waves. At low beam densities, the inequality $|\varpi - k_\parallel v_b + \Omega_0|\leq\eta\,\Omega_0$ can be marginally fulfilled only by waves that match Bell's semi-MHD approximation of $|\varpi|\ll\Omega_0$. The growth rate (\ref{eqnGBell}) describes these waves well in an intermediate range of densities, but overestimates their growth if one assumes a cosmic-ray beam of very low density and temperature.

For beam currents larger than $j_\mathrm{crit} \approx 2\,e\,n_0\,v_A$, however, waves which match the condition for exact gyroresonance, $|\varpi - k_\parallel v_b + \Omega_0|=0$ become unstable. In this regime, the semi-MHD approximation is inapplicable. A more exact analysis reveals that two modes emerge as the current increases, both of which are left-hand polarized in the background frame: a long-wavelength shear-Alfv\'en wave and a short-wavelength ion-cyclotron wave. Their growth rate has an upper bound at $\gamma\approx\Omega_0$.

In addition to these two background-gyroresonant modes, a third mode exists which is additionally gyroresonant with the electron population. Assuming a cold current-compensating electron population, this third mode grows generally much more slowly, but it merges with the ion-cyclotron mode at large current strengths.

The relevance of the nearly gyroresonant character of Bell's instability is that it clearly identifies the driving source of free energy. Since the gyrofrequency of the background ions is close to the frequency of the shear-Alfv\'en waves that are excited, gyroresonant acceleration and scattering will tend to isotropize the background ions in the wave frame and thus reduce the free energy in the system. At the same time, the fast magnetosonic waves driven by the streaming instability will act as a symmetric counterpart and isotropize the CR beam. Eventually, the relative drift between the background ions and the beam will be too small to excite further wave growth. Reaching this saturation stage will take far longer in the case of near-gyroresonance than for exact gyroresonance, but that is precisely why these modes grow so much more slowly in the low-density Bell regime.

\acknowledgments
M.~S.~W. is supported by the Deutsche~Forschungsgemeinschaft and the Defense~Threat~Reduction Agency and thanks George Morales for many fruitful discussions. Several calculations were performed at NERSC, a U.S. Department of Energy Office of Science User Facility (Contract No. DE-AC02-05CH11231).

\bibliographystyle{aasjournal}
%\bibliography{BellRefs}

\end{document}